\documentclass{IEEEtran}
\usepackage{amsmath,amssymb,amsfonts}
\usepackage{graphicx}
\usepackage{textcomp,nicefrac}
\usepackage{xurl}
\usepackage{doi}

\usepackage{tabularx}
\usepackage{booktabs}
\usepackage{multirow}

\usepackage{listings}

\usepackage{fancyvrb}
\usepackage{relsize}

\usepackage[acronym, nohypertypes={acronym,notation}]{glossaries}

\newacronym{hl}{HL}{High-Luminosity}
\newacronym{lhc}{LHC}{Large Hadron Collider}
\newacronym{soc}{SoC}{System-on-Chip}
\newacronym{zusp}{ZUS+}{Zynq UltraScale+}
\newacronym{pxe}{PXE}{Preboot Execution Environment}
\newacronym{fsbl}{FSBL}{First Stage Bootloader}
\newacronym{mpsoc}{MPSoC}{Multi-Processor System-on-Chip}
\newacronym{pl}{PL}{Programmable Logic}
\newacronym{ps}{PS}{Processing System}
\newacronym{apu}{APU}{Application Processing Unit}
\newacronym{rpu}{RPU}{Real-Time Processing Unit}
\newacronym{pmu}{PMU}{Platform Management Unit}
\newacronym{csu}{CSU}{Configuration Security Unit}
\newacronym{pcw}{PCW}{PS Configuration Wizard}
\newacronym{qspi}{QSPI}{Quad Serial Peripheral Interface}
\newacronym{uart}{UART}{Universal Asynchronous Receiver-Transmitter}
\newacronym{fpga}{FPGA}{Field-Programmable Gate Array}
\newacronym{atf}{ATF}{Arm Trusted Firmware}
\newacronym{bios}{BIOS}{Basic Input/Output System}
\newacronym{tftp}{TFTP}{Trivial File Transfer Protocol}
\newacronym{smc}{SMC}{Secure Monitor Call}
\newacronym{nfs}{NFS}{Network File System}
\newacronym{rootfs}{rootfs}{root file system}
\newacronym{xsa}{XSA}{Xilinx Support Archive}
\newacronym{xml}{XML}{Extensible Markup Language}
\newacronym{ascii}{ASCII}{American Standard Code for Information Interchange}
\newacronym{som}{SoM}{System-on-Module}
\newacronym{rgmii}{RGMII}{Reduced Gigabit Media-Independent Interface}
\newacronym{sgmii}{SGMII}{Serial Gigabit Media-Independent Interface}
\newacronym{sata}{SATA}{Serial AT Attachment}
\newacronym{ssd}{SSD}{Solid-State Drive}
\newacronym{mio}{MIO}{Multiplexed Input/Output}
\newacronym{axi}{AXI}{Advanced eXtensible Interface}
\newacronym{bram}{BRAM}{Block RAM}
\newacronym{ip}{IP}{Intellectual Property}
\newacronym{serdes}{SerDes}{Serializer/Deserializer}
\newacronym{fmcp}{FMC+}{FPGA Mezzanine Card Plus}
\newacronym{pll}{PLL}{Phase-Locked Loop}
\newacronym{ppu}{PPU}{Peripheral Protection Unit}

\def\BibTeX{{\rm B\kern-.05em{\sc i\kern-.025em b}\kern-.08emT\kern-.1667em\lower.7ex\hbox{E}\kern-.125emX}}

\begin{document}
\title{Split Boot - True Network-Based Booting on Heterogeneous MPSoCs}
\author{Marvin Fuchs, Luis E. Ardila-Perez, Torben Mehner, and Oliver Sander
\thanks{Manuscript submitted September 10, 2022; revised November 11, 2022. This research acknowledges the support by the Doctoral School
\emph{``Karlsruhe School of Elementary and Astroparticle Physics:
Science and Technology''}

M. Fuchs (corresponding author, email: marvin.fuchs at kit.edu),
L. E. Ardila-Perez, T. Mehner and O. Sander are with the Institute for
Data Processing and Electronics (IPE) of the Karlsruhe Institute of Technology,
Hermann-von-Helmholtz-Platz 1, D-76344 Eggenstein-Leopoldshafen, Germany}}

% ORCID (I recommend you create a profile)
% Marvin Fuchs          https://orcid.org/0000-0002-4146-5846
% Luis E. Ardila-Perez  https://orcid.org/0000-0002-7485-8267
% Torben Mehner         https://orcid.org/0000-0002-8506-5510
% Oliver Sander         https://orcid.org/0000-0002-0959-4744

\maketitle

\begin{abstract}

In the context of the \acrfull{hl} upgrade of the \acrshort{lhc}, many custom ATCA electronics boards are being designed containing heterogeneous \acrfull{soc} devices, more specifically the Xilinx \acrfull{zusp} family. While the application varies greatly, these devices are regularly used for performing board management tasks, making them a fundamental element in the correct operation of the board. The large number of hundreds of \acrshort{soc} devices creates significant challenges in their firmware deployment, maintenance, and accessibility. 

Even though U-Boot on \acrshort{zusp} devices supports network boot through the \acrfull{pxe}, the standard \acrshort{zusp} boot process contains application-specific information at earlier boot steps, particularly within the \acrfull{fsbl}. This prevents the initialization of several devices from a universal image. Inspired by the \acrshort{pxe} boot process on desktop PCs, this paper describes split boot, a novel boot method tailored to the specific needs of the \acrshort{zusp}. All application-specific configuration is moved to a network storage device and automatically fetched during the boot process. We considered the entire process, from firmware and software development to binary distribution in a large-scale system. The developed method nicely integrates with the standard Xilinx development toolset workflow.

\end{abstract}

\begin{IEEEkeywords}
Booting, Large-Scale Experiments, MPSoC, Network Booting, PXE, System-on-Chip, Zynq Ultrascale+
\end{IEEEkeywords}

\section{Introduction}
\label{sec:introduction}

\begin{figure}[b!]
    \centering
    \includegraphics[width=3.4in]{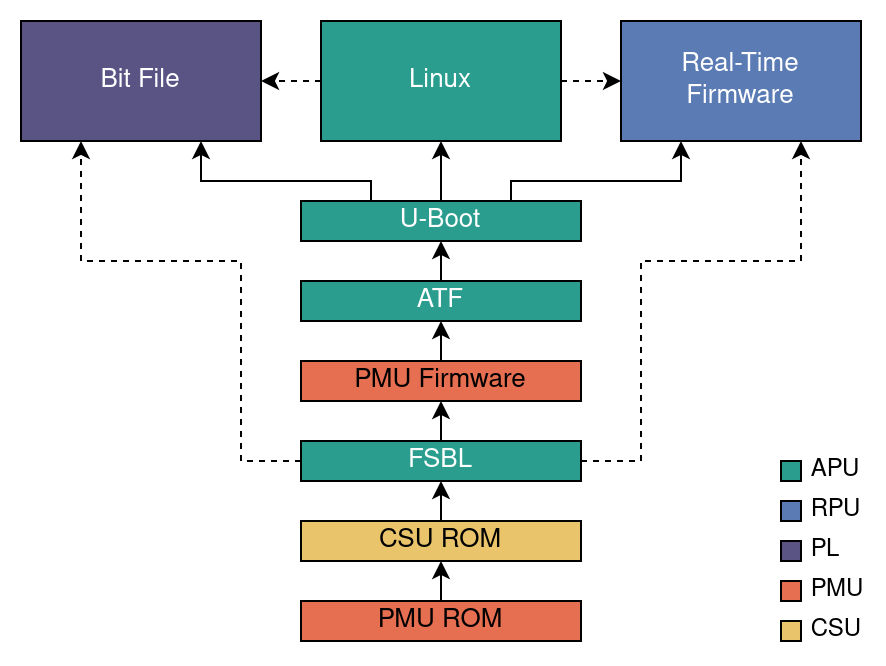}
    \caption{Default software stack used to boot Xilinx Zynq UltraScale+ devices. The solid lines describe one possible example of a boot process, whereas the dashed lines show alternative possibilities to load a bitfile for the PL and the firmware for the \acrshort{rpu}.}
    \label{fig:defaultBootorder}
\end{figure}

\IEEEPARstart{T}{he} Xilinx \acrfull{zusp} devices are heterogeneous \acrlong{mpsoc}s (\acrshort{mpsoc}s) that, in addition to the \acrfull{pl}, contain a \acrfull{ps} with a number of hard processing units, such as an ARM Cortex-A53 named \acrfull{apu}, an ARM Cortex-R5 named \acrfull{rpu}, and the \acrfull{pmu} based on the MicroBlaze architecture\,\cite{xlnx_ug1137}. Even though not all processors have to be involved in the boot process, it usually relies on several of them. To make the \acrshort{zusp} devices deployable in a wide range of applications, they are designed to be highly configurable. To a certain extent, this also applies to the boot process, as shown in Fig.~\ref{fig:defaultBootorder}. For example, it is possible to load both, the bitfile for the \acrshort{pl} and the firmware for the \acrshort{rpu}, in either the \acrfull{fsbl}, the second-stage boot loader U-Boot or from Linux. In some cases it is also possible to change the order, for example to load the \acrshort{pmu} firmware either before or after the \acrshort{fsbl}.

The first stage in the boot process that contains application-specific information is the \acrshort{fsbl}. Vivado generates C code to configure the \acrshort{ps} and to optionally divide it in multiple subsystems via an isolation-configuration according to the settings selected in the Vivado \acrfull{pcw} GUI\,\cite{pcw_video}. When the \acrshort{fsbl} software project is set up, this source code is automatically integrated. 

The \acrshort{fsbl} contains information on how to configure the various internal clocks, the interfaces to the \acrshort{pl}, and external interfaces such as the \acrshort{uart}s or the network. The application-specific code contained in it to do this, is one of the reasons why the \acrshort{fsbl} cannot be factory-saved to a non-volatile memory within the \acrshort{mpsoc}. Thus, it is one of the first components loaded from an external storage (e.g. a \acrshort{qspi} memory or an SD Card). The same storage location is usually also used for the \acrshort{pmu} firmware, the \acrfull{atf}, and U-Boot. In contrast, entirely generic software like the \acrshort{pmu} ROM and the \acrfull{csu} ROM is stored on non-volatile memory within the \acrshort{zusp}\,\cite{xlnx_ug1085}.

The goal of the modified boot process presented in this paper is to fetch all application-specific data including the \acrshort{ps} configuration from a network source. This is, however, not trivial because the \acrshort{fsbl} itself is a low-level stage in the boot process and it is not designated to communicate via a network connection. Just as with PCs that boot via \acrfull{pxe}, the greatest advantage for \acrshort{mpsoc}s that obtain all application-specific data from the network appears in large systems where many devices need to be maintained. One example is the distribution of updates in a large and distributed system, comprised of many identically configured boards. In its most efficient implementation, split boot enables accomplishing the task by only updating the single network storage and rebooting the devices. Significant time is saved compared to having to flash the local storage of each board. In the remainder of this contribution, we present a two-step approach that essentially supports fetching the entire \acrshort{ps} configuration via network and applying it to the \acrshort{ps}.

\section{Related Work}

U-Boot already features \acrshort{pxe}, which allows devices to boot into an operating system such as Linux via the network. However, it does not cover the configuration of the \acrshort{ps}\,\cite{xlnx_ug1144}. Such a functionality is not provided, because \acrshort{pxe} was originally designed for computer networks rather than networks of highly configurable \acrlong{soc}s (\acrshort{soc}s) such as the \acrshort{zusp} family. Modifications to the \acrshort{fsbl} on \acrshort{mpsoc} devices might be required on a regular basis due to containing application-specific information. This is a major drawback compared to the boot of a desktop PC, where updates to the \acrshort{bios} and all other software used before the second-stage boot loader are very rarely necessary. Research regarding \acrshort{pxe} usually targets desktop PCs\,\cite{5691309} or servers\,\cite{9339366}, but not embedded devices. Xilinx provides means to adapt the boot process based on the application domain\,\cite{xlnx_ug1137} and further describes in a patent the boot process possibilities of \acrshort{mpsoc} devices\,\cite{xil_pat_boot}. However, the ability to load the \acrshort{ps} configuration from a remote location is not mentioned. In the context of the \acrfull{hl} upgrade of the \acrshort{lhc}, active research is being conducted about the boot process of \acrshort{mpsoc} devices. To date, though, work has focused primarily on investigating and securing the possibilities provided by Xilinx\,\cite{cern_dzemaili} and building the Linux distribution for use on the device\,\cite{cern_mor}.

\section{Split Configuration Approach}
\label{sec:splitconfigapproach}

Simply moving the entire configuration of the \acrshort{ps} part of a \acrshort{zusp} \acrshort{mpsoc} to a network storage is not feasible. Some initial configuration is needed to bring up the essential functionality of the device. This includes, first and foremost, the network interface and the configuration of the DDR memory controller, but also some internal configuration. While this configuration is board-specific, it is not application-specific. As a conclusion, we propose using a base configuration which is static and reduced to the absolute minimum to boot into U-Boot with network access. This approach is similar in many ways to that of a PC \acrshort{bios}. Updates to the base firmware are possible, but they are expected to happen rarely.

\begin{figure}[b!]
    \centering
    \includegraphics[width=3.4in]{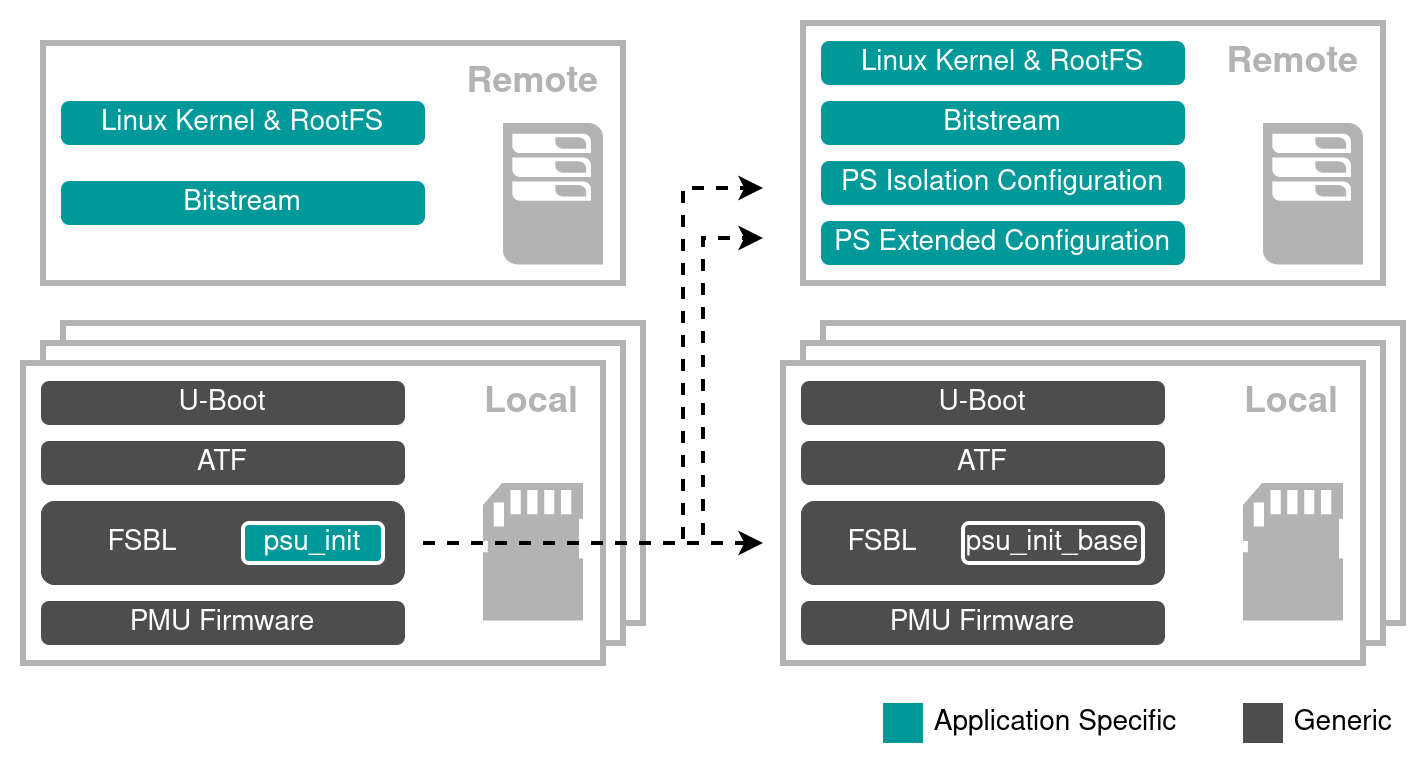}
    \caption{The split boot approach removes from the \acrshort{fsbl} the application-specific \acrshort{ps} configuration and the isolation configuration. As a result, only generic information remains on the local boot medium.}
    \label{fig:datastorage}
\end{figure}

The application-specific data for the \acrshort{ps} can be split into two tasks. The first one includes the configuration of all the individual components like clocks within the \acrshort{ps}, the \acrshort{ps}-\acrshort{pl} interfaces, and some peripherals like the DDR memory. The second task is the application of the so-called isolation-configuration, which divides the \acrshort{ps} into multiple subsystems and defines access permissions between them. For both, we propose to move the application-specific data into separate binary configuration files, which are then fetched from a network source and applied by U-Boot during the boot process. Fig.~\ref{fig:datastorage} shows how removing the application-specific configuration data from the \acrshort{fsbl} leads to a system where only generic software remains on the local boot medium and everything application-specific is stored on the network.

\section{The Modified Boot Process}
\label{sec:modbootprocess}

\begin{figure*}[tb!]
    \centering
    \includegraphics[width=\textwidth]{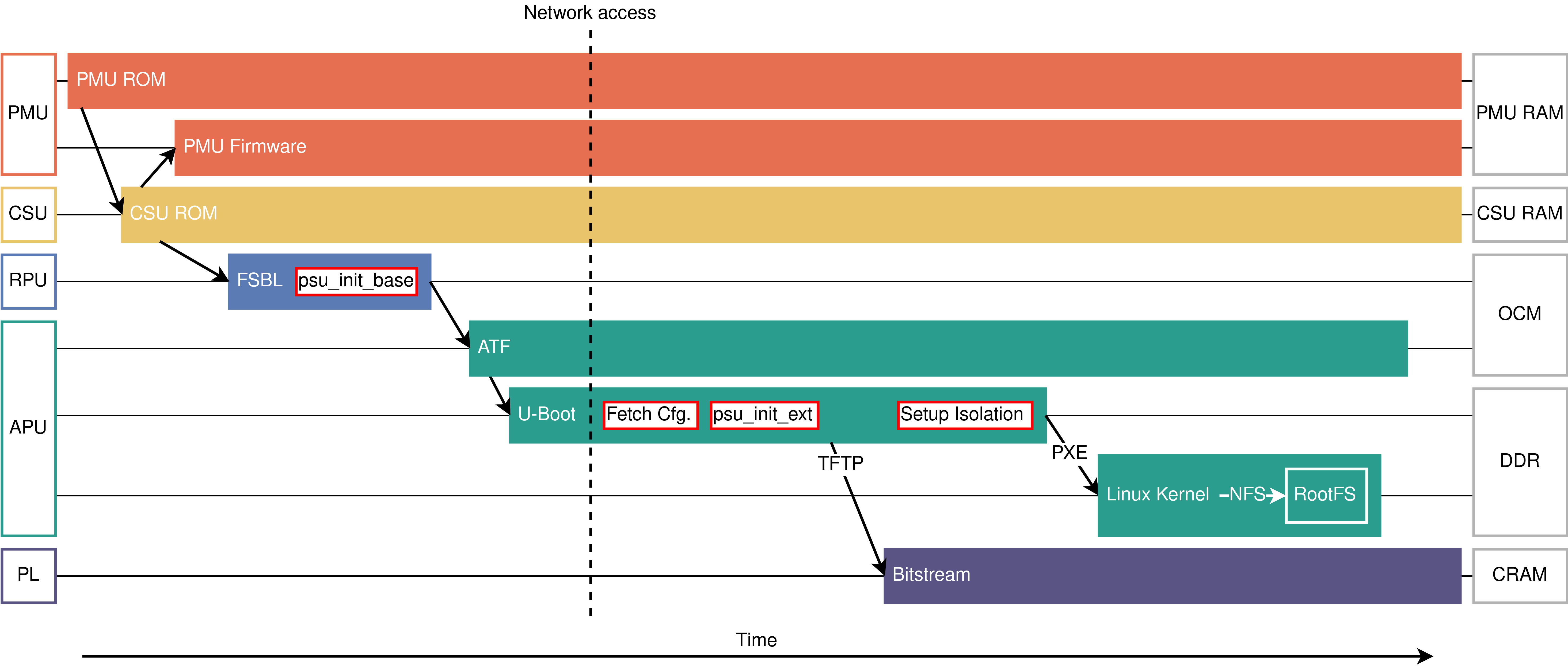}
    \caption{Modified boot process for Xilinx \acrlong{zusp} devices. The modifications are represented by the red bordered boxes. The first modification in the boot process, \textit{psu\_init\_base}, marks were usually the configuration of the \acrshort{ps} would happen. Here, the source code to completely configure the \acrshort{ps} is replaced by source code that applies a base configuration, allowing to continue the boot process and using a network interface later on. The remaining modifications can all be found in U-Boot. They include fetching configuration data, modifying the configuration of the \acrshort{ps} (\textit{psu\_init\_ext}), and optionally configuring the isolation within the \acrshort{ps}.}
    \label{fig:bootflow}
\end{figure*}

Xilinx \acrshort{zusp} devices require multiple tweaks in the default boot process to allow for changes to the \acrshort{ps} configuration after it has been initially configured in the \acrshort{fsbl}. An overview of these modifications is provided by the example boot sequence in Fig.~\ref{fig:bootflow}, where the changes are shown as white boxes with red frames. The boot procedure starts as usual with the software components \acrshort{pmu} ROM and \acrshort{csu} ROM stored on non-volatile memory within the \acrshort{zusp}. Afterwards, the \acrshort{pmu} firmware is started, in this case, before the \acrshort{fsbl}. 

The \acrshort{fsbl} contains the first modification in the proposed boot process. Usually, at this stage, the \acrshort{ps} gets initialized with its complete configuration. During the split boot process, however, this is where the \acrshort{ps} receives its base configuration (psu\_init\_base), which includes the boot related peripheral and memory configuration.

The source code that is used to configure the \acrshort{ps} is generated by Vivado according to the options selected in the \acrshort{pcw} and inserted into the \acrshort{fsbl} automatically. Because of this, the software architecture of the \acrshort{fsbl} allows us to easily replace this part of the source code, for example with that of our generic base configuration.

When the \acrshort{fsbl} has finished its execution, it hands over to the \acrshort{atf}. The \acrshort{atf} is a reference implementation of an ARM secure world software provided by Xilinx and in the example depicted in Fig.~\ref{fig:bootflow} the first software executed on the \acrshort{apu}. Such a software is necessary to utilize the Armv8-A Exception Model that is implemented in the \acrshort{apu}\,\cite{arm_exc_lvl}.

When the \acrshort{atf} is running, U-Boot can be started. It contains the remaining modifications to the original boot process. The first modification (Fetch Cfg.) loads all the required configuration data using standard U-Boot features from a \acrshort{tftp} server. This includes, in addition to the regular Linux kernel, a custom binary file for the modification of the \acrshort{ps} configuration, a file to configure the isolation within the \acrshort{ps}, and the \acrshort{pl} bitstream. The next step is to extend the configuration of the \acrshort{ps} to its complete state (psu\_init\_ext). U-Boot running on ARM Exception Level 2 is not allowed to access the required configuration registers directly. Instead, it is possible to request the \acrshort{atf} running at Exception Level 3 via a \acrfull{smc} to instruct the \acrshort{pmu} firmware to access a configuration register. The \acrshort{pmu} has unrestricted access to all configuration registers within the \acrshort{ps}\,\cite{xlnx_ug1085}. Using this methodology, the \acrshort{atf} can remain unmodified, while only a minor modification to the \acrshort{pmu} firmware is required to temporarily allow U-Boot to make all these requests until the \acrshort{ps} is fully configured.

After the reconfiguration of the \acrshort{ps} is finished, it might be necessary to rebind some U-Boot drivers, for example the one used for Ethernet. Then, the bitstream can be written to the \acrshort{pl}. It is also possible to configure an isolation within the \acrshort{ps} (Setup isolation) if desired. At this point, the system is fully configured, operational, and behaves exactly the same as it would with the traditional boot.
Finally, U-Boot can continue to boot Linux on the \acrshort{apu}. In the example boot process shown in Fig.~\ref{fig:bootflow}, \acrshort{pxe} is used to load the Linux kernel, and the kernel itself uses \acrshort{nfs} to mount the \acrshort{rootfs}.  

Modifying the configuration of some critical components within the \acrshort{ps} is not possible from U-Boot. This applies to the DDR interface and a very limited number of other configuration registers as shown in Table~\ref{tab:notAllowedReg}. However, for many resources that are used by U-Boot but are not essential for it to run, it is possible to overwrite the configuration. Activating the isolation within the \acrshort{ps} from U-Boot brings some limitations as well. All software that is running while the isolation is being activated must observe the restrictions enforced by it. If the isolation includes a restriction of the memory ranges used by the \acrshort{apu} for instance, it is mandatory that U-Boot observes this restriction before the activation of the isolation. Otherwise U-Boot might lose access to essential data when the isolation is activated, which can lead to undesired behaviour or even to a crash of the system.

\begin{table}[b]
    \caption{List of Registers That Cannot Be Modified From U-Boot\,\cite{kit_fuchs}.}
    \label{tab:notAllowedReg}
    \begin{center}
    \resizebox{3.5in}{!}{
    \begin{tabularx}{3.5in}{llllX}
    \toprule
    \textbf{Register}   & \textbf{Mask}    & \textbf{Reason}\\
    \midrule
    \verb|0xFD1A0030|   & \verb|0xFE7FEDEF| & \multirow{7}{1.65in}{Part of the configuration of the DDR.}\\
    \verb|0xFD1A002C|   & \verb|0x00717F00| &\\
    \verb|0xFD1A002C|   & \verb|0x00000008| &\\
    \verb|0xFD1A002C|   & \verb|0x00000001| &\\
    \verb|0xFD1A0044|   & \verb|0x00000002| &\\
    \verb|0xFD1A004C|   & \verb|0x00003F00| &\\
    \verb|0xFD1A0080|   & \verb|0x00003F07| &\\
    \midrule
    \verb|0xFF260020|   & \verb|0xFFFFFFFF| & \multirow{2}{1.65in}{Can be read from the \acrshort{atf}.}\\
    \verb|0xFF260000|   & \verb|0x00000001| &\\
    \midrule
    \verb|0xFD0C00AC|   & \verb|0xFFFFFFFF| & \multirow{4}{1.65in}{\acrshort{sata} Port Phy configuration registers initialized with reset values\,\cite{register_reference}. Modifying these registers causes a crash.}\\
    \verb|0xFD0C00B0|   & \verb|0xFFFFFFFF| &\\
    \verb|0xFD0C00B4|   & \verb|0xFFFFFFFF| &\\
    \verb|0xFD0C00B8|   & \verb|0xFFFFFFFF| &\\        
    \bottomrule
    \end{tabularx}
    }
    \end{center}
\end{table}

\begin{table}[tb]
    \caption{List of Functions Represented in the Binary Configuration Files.}
    \label{tab:funcsInBin}
    \begin{center}
    \begin{tabularx}{3.5in}{lX}
    \toprule
    \textbf{Name}   & \textbf{Action}\\
    \midrule
    \verb|PSU_Mask_Write|               &   Read-modify-write\\
    \addlinespace
    \verb|mask_poll|                    &   Polls until a 1 occurs in the masked part of the register or a specified number of attempts in exceeded\\
    \addlinespace
    \verb|mask_pollOnValue|             &   Polls until the masked part of the register matches the desired value, or until a certain number of attempts is exceeded\\
    \addlinespace
    \verb|mask_delay|                   &   Delay for a specified duration\\
    \addlinespace
    \verb|serdes_illcalib|              &   Calibration algorithm for \acrshort{serdes}\\
    \addlinespace
    \verb|serdes_fixcal_code|           &   Calibration algorithm\\
    \addlinespace
    \verb|serdes_enb_coarse_saturation| &   Activates the coarse saturation logic for \acrshort{pll}s of all four GT lanes of the \acrshort{mpsoc}\\
    \addlinespace
    \verb|psu_init_xppu_aper_ram|       &   Initialization of the \acrshort{ppu}\\
    \bottomrule
    \end{tabularx}
    \end{center}
\end{table}

\section{Custom Configuration Files}
\label{sec:customconfigfiles}

The PS configuration files that are stored and later fetched from the network are encoded using a binary format. This is done to efficiently process them in U-Boot. Despite the many features of U-Boot, it is still low-level software. Therefore, working with more complex data formats based on \acrshort{ascii} like \acrshort{xml} would significantly increase the overhead. While human readable code would be an advantage, the wish to manually change about one thousand 32-bit registers is rather unlikely.

The internal structure of the binary configuration files is strongly inspired by the architecture of the source code that is used in the \acrshort{fsbl} to configure the \acrshort{ps}. This code can be unwrapped to a long list of calls of eight different functions listed in Table~\ref{tab:funcsInBin}\,\cite{kit_fuchs}. The configuration of the \acrshort{ps} is therefore fully represented by this list of function calls including the respective call arguments. Therefore, this is the only information that must be stored in the binary configuration files. Listing~\ref{lst:psuCS} and Listing~\ref{lst:psuCSBin} show the encoding of the function calls in the binary format. All call arguments of the functions in Table~\ref{tab:funcsInBin} are 32-bit values, which can also be realized by macros in the source code. It is possible to represent each of the eight different functions with an unique 32-bit ID. As seen in Listing~\ref{lst:psuCS} and Listing~\ref{lst:psuCSBin} the function \verb|PSU_Mask_Write| has the ID \verb|0x00000001|. The binary file is now composed of the list of function calls from the \acrshort{fsbl} encoded in this format. It is possible to navigate through the different function calls in such a binary file because the number of arguments of each function is constant and known. Finally, a distinct unique ID \verb|0x0000000F| is used to mark the end of the file, as can be seen at the end of the file in Listing~\ref{lst:psuCSBin}.

\begin{figure}[t]
\begin{lstlisting}[language=C, breaklines, basicstyle=\relsize{-1}\ttfamily, label=lst:psuCS, captionpos=b, caption=Source code snippet from psu\_init.c.]
PSU_Mask_Write(CRL_APB_RPLL_CFG_OFFSET,
               0xFE7FEDEFU, 0x7E4B0C82U);
PSU_Mask_Write(CRL_APB_RPLL_CTRL_OFFSET,
               0x00717F00U, 0x00015400U);
PSU_Mask_Write(CRL_APB_RPLL_CTRL_OFFSET,
               0x00000008U, 0x00000008U);
PSU_Mask_Write(CRL_APB_RPLL_CTRL_OFFSET,
               0x00000001U, 0x00000001U);
PSU_Mask_Write(CRL_APB_RPLL_CTRL_OFFSET,
               0x00000001U, 0x00000000U);
mask_poll(CRL_APB_PLL_STATUS_OFFSET,
          0x00000002U);
\end{lstlisting}
\end{figure}

\begin{figure}[t]
\begin{lstlisting}[language=Clean, breaklines, basicstyle=\relsize{-1}\ttfamily, label=lst:psuCSBin, captionpos=b, caption=Encoding of the source code in Listing~\ref{lst:psuCS} in a binary configuration file.]
00000001 FF5E0034 FE7FEDEF 7E4B0C82
00000001 FF5E0030 00717F00 00015400
00000001 FF5E0030 00000008 00000008
00000001 FF5E0030 00000001 00000001
00000001 FF5E0030 00000001 00000000
00000003 FF5E0040 00000002 0000000F
\end{lstlisting}
\end{figure}

\section{PSU Configuration Generator}
\label{sec:psuconfiggen}

A Python tool called the PSU Configuration Generator was developed to keep the effort of developing a project using the split boot mechanism to a minimum. This tool handles, among other things, the generation of the binary configuration files. It was designed to integrate seamlessly with the development tools provided by Xilinx. Thus, the *.xsa (\acrlong{xsa}) files exported from Vivado are used as input data. Within this archive, \textit{psu\_init.c} and \textit{psu\_init.h} contain the C source code which is used in the \acrshort{fsbl} to configure the \acrshort{ps}. They also contain a more abstract description of the configuration of the \acrshort{ps} and the \acrshort{pl} in the \acrshort{xml} file \textit{zusys.hwh}. The current implementation of the PSU Configuration Generator uses the *.xsa file containing the complete \acrshort{ps} configuration as input. 

The files \textit{psu\_init.c} and \textit{psu\_init.h} are parsed to identify the function calls that need to be written to the configuration files. The parser uses a depth-first search to find all calls of the eight functions listed in Table~\ref{tab:funcsInBin}, starting by the functions that can be called directly from the \acrshort{fsbl}. If a function call is allowed in U-Boot, or in other words, if the addressed registers can be modified from U-Boot, the function call is encoded in the binary format, resolving all macros contained in it, and appended to the binary output file. Furthermore, to be adaptable, the PSU Configuration Generator allows skipping selected registers or ignoring some function calls specified in a JSON file. There are only a few interfaces to the \acrshort{fsbl} that represent root nodes. Using the functions in Table~\ref{tab:funcsInBin}, all nodes representing termination conditions for the depth-first search are identified. Both, the root and termination nodes form the boundary conditions for the search algorithm.

The file \textit{psu\_init.c} contains two main interfaces to the \acrshort{fsbl}. One to configure the \acrshort{ps} and one to set up the isolation. To execute these two actions separately from U-Boot, the PSU Configuration Generator offers the possibility to export separate configuration files. Additionally, it provides the option to extract the bitfile for the \acrshort{pl} from the *.xsa archive. Therefore, achieving a higher degree of automation as all these files need to be copied to the same \acrshort{tftp} server.

\section{Development Workflow}
\label{sec:workflow}

To make split boot usable in real world applications, it is important to integrate the required modifications to the different software components and the additional steps in the development process with the default workflow of the Xilinx development tools. To achieve this, an approach based on two Vivado projects was chosen. One project represents the base configuration of the \acrshort{ps}, and optionally also of the \acrshort{pl}, which are applied at the \acrshort{fsbl}. The other project contains the complete configuration, which is fetched by U-Boot via the network. Both Vivado projects are integrated into a workflow that is divided into two independent sub-processes. The first one leads to the generation of all files that are needed in the boot routine before network access is possible. The second sub-process produces the necessary files that can be loaded from the network. This distinct separation enables only the second sub-process to be required for each new project. The first sub-process, containing only generic data, is only executed once per hardware platform. As a result, the development effort with split boot is comparable to a project without it.

\subsection{Creation of the Base Configuration}

\begin{figure}[tb!]
    \centering
    \includegraphics[width=2.0in]{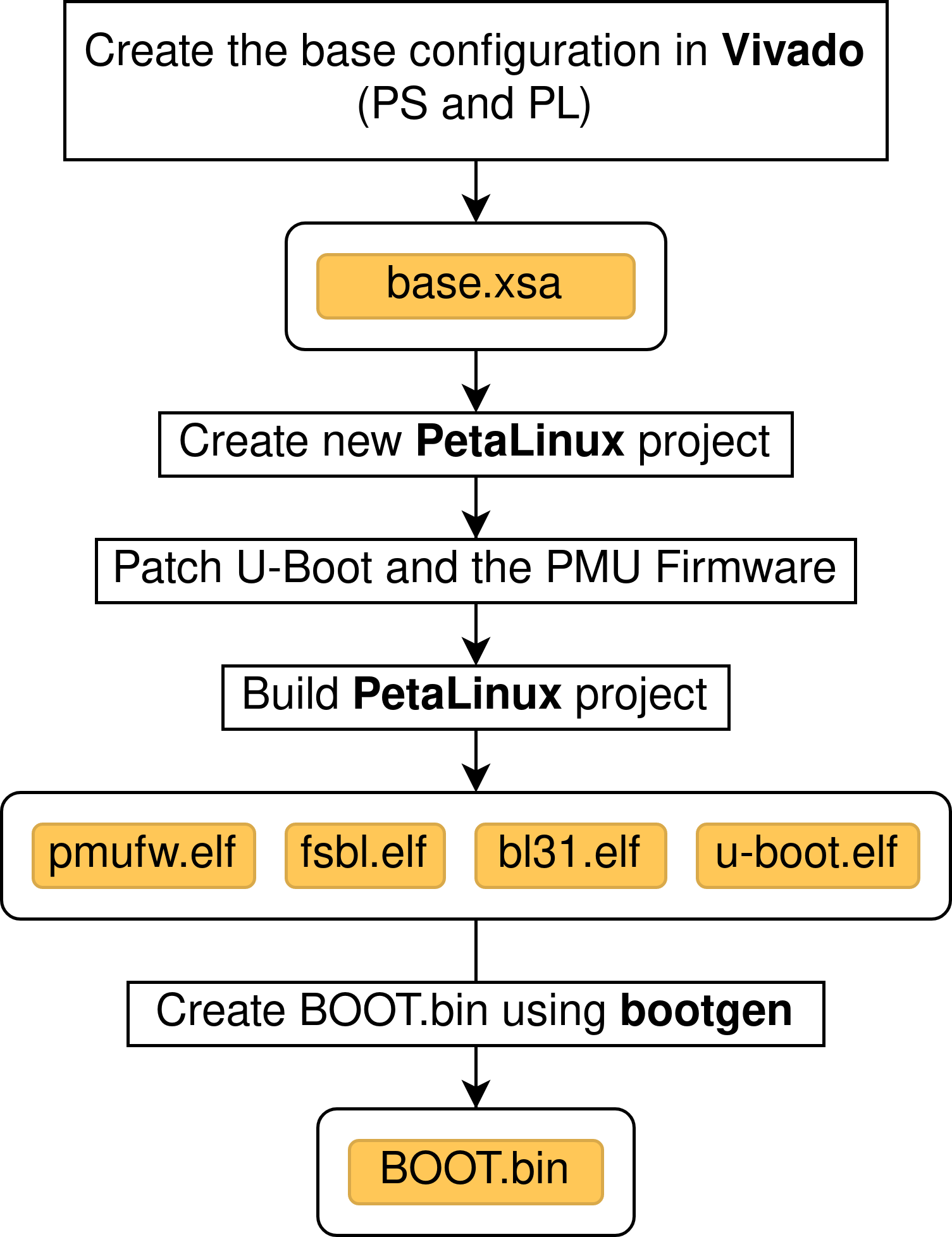}
    \caption{Creation flow of all software and firmware components to be stored on the local boot medium. They are based on a Vivado project representing the base configuration of the \acrshort{ps} and optionally also the \acrshort{pl}. Output files are packed in the single boot image called \textit{BOOT.bin}.}
    \label{fig:createfilesbase}
\end{figure}

The creation of all files needed for the early stages of the boot process, before a network connection can be utilized, are depicted in Fig.~\ref{fig:createfilesbase}. In particular, they also contain the base configuration for the \acrshort{ps}. As can be seen, these files are entirely based on the *.xsa file exported from the Vivado project representing the base configuration \textit{base.xsa}. The PetaLinux tools are the only additional tools from the Xilinx development suite that are required in the process. After creating a respective PetaLinux project, the tools automate the process of building all individual software components. However, one manual step might be required if the complete configuration includes an isolation setting because the memory regions used by U-Boot need to be restricted according to it. This can be achieved in the device tree, and it is the only limitation we have thus far observed as a result of the activation of the isolation within the \acrshort{ps} from U-Boot. As can be seen in Fig.~\ref{fig:createfilesbase}, patches are used to apply the required modifications to U-Boot and the \acrshort{pmu} firmware. The use of patch files is an integral part of developing with the PetaLinux software suite, and thus both the creation and the application during the build process is automated.

The patch applied to the U-Boot source code is used to add the functionality to modify the configuration of the \acrshort{ps} and to apply the isolation. This functionality is packaged in the custom U-Boot-command \textit{psuinite}. As an argument this command needs the address of the configuration file to be applied in memory. It then iterates over the configuration file and executes the function calls listed there. The command contains implementations of all functions listed in Table~\ref{tab:funcsInBin}. The source code is derived from the implementations in \textit{psu\_init.c} and only slightly modified to use the drivers available in U-Boot and to request access to the required configuration registers via \acrshort{smc} from the \acrshort{pmu} firmware instead of accessing them directly. A flag can be passed to the command if the access to the configuration registers from the \acrshort{apu} should be locked in the \acrshort{pmu} firmware after the configuration file is applied. Finally, a debugging flag exists that enables print outputs for each register access made. Another patch applied to the U-Boot source code inserts all the additional steps required by split boot to the regular steps U-Boot performs to boot the system. This patch also includes checks if the additional steps were executed successfully or not. If a failure is detected, the boot process is immediately aborted with an error message because the errorless execution of each of these steps is essential for a successful boot.

The patch applied to the \acrshort{pmu} firmware is required to allow U-Boot to access all needed configuration registers via \acrshort{smc} calls. By default, the \acrshort{pmu} firmware verifies that the requesting instance is authorized to access the requested resource. This mechanism must be temporarily disabled until U-Boot has completed all required configuration register accesses. The patch enables all such accesses from the start of the \acrshort{pmu} firmware and gives U-Boot the option to restore the default access control when the configuration has been extended to its complete state.

PetaLinux tools are able to build the \acrshort{pmu} firmware, the \acrshort{fsbl}, the \acrshort{atf}, and U-Boot once the patches have been applied. Since this \acrshort{fsbl} is based on the \textit{base.xsa}, it already contains the desired configuration for the \acrshort{ps} in \textit{psu\_init.c} and \textit{psu\_init.h}, so modifying these files is not necessary. However, the \acrshort{fsbl} contains one more application-specific section. The structure \textit{XPm\_ConfigObject} contains, among other things, the information about which components in the \acrshort{ps} will be used in the given configuration. One of the final steps in the \acrshort{fsbl} is to send this information to the \acrshort{pmu} firmware. If a component is not marked as active in this structure, it is not possible to activate it later purely via configuration registers. One workaround for this limitation is to mark every node in the \textit{XPm\_ConfigObject} as active. The downside is that this increases the power consumption of the \acrshort{mpsoc} as all nodes will be powered, which also leads to potential security vulnerabilities. Therefore, it is still being investigated if this structure can be modified at a later stage of the boot process. 

Having the \acrshort{pmu} firmware, the \acrshort{fsbl}, the \acrshort{atf}, and U-Boot ready, the final step is to use the Xilinx tool \textit{bootgen} to package them in a boot image. This file can then be copied to a local boot medium such as an SD Card.

\subsection{Creation of the Complete Configuration}

\begin{figure}[tb!]
    \centering
    \includegraphics[width=3.4in]{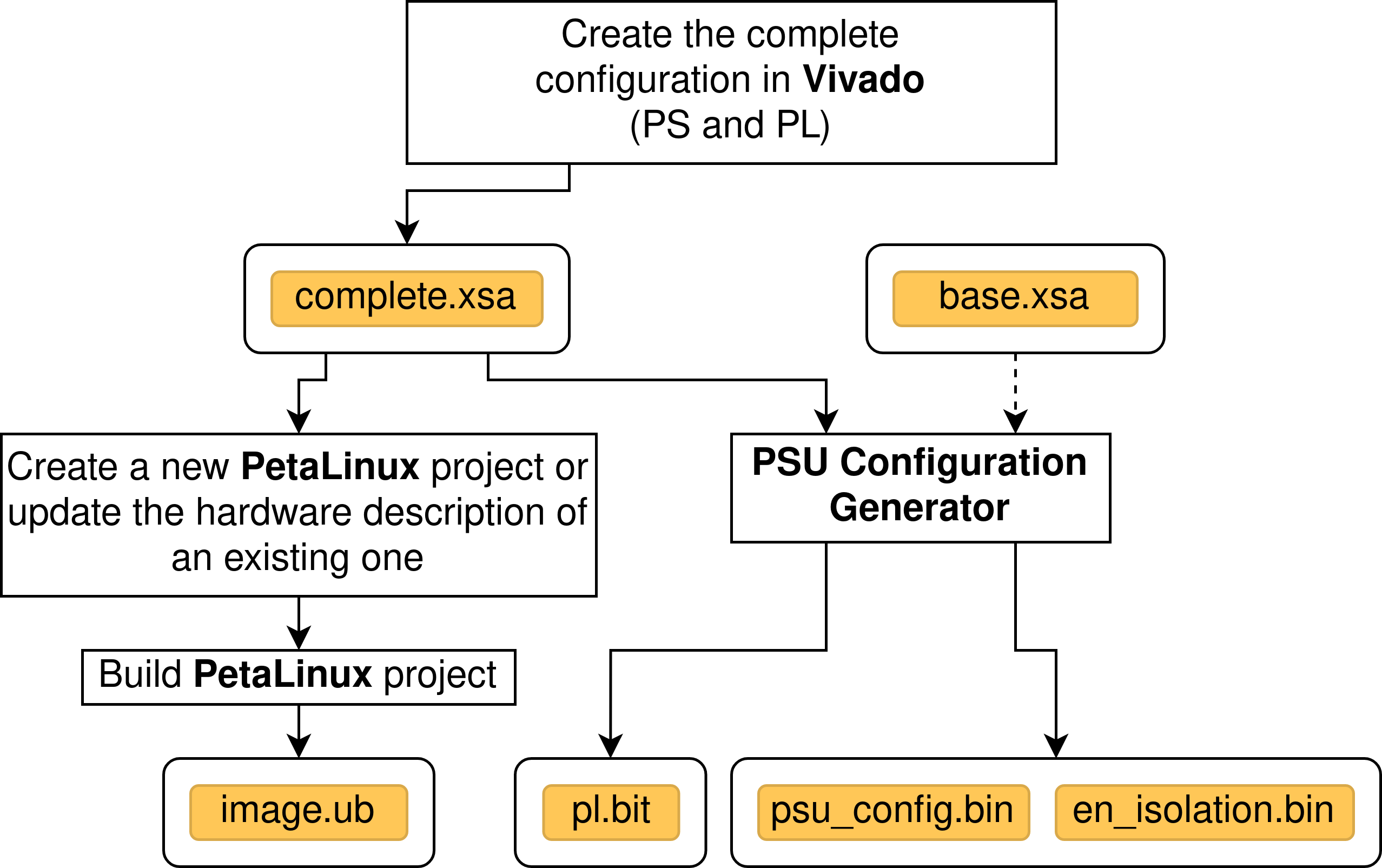}
    \caption{Creation of all software and firmware components that are fetched from the network during boot. They are based on a Vivado project representing the complete configuration of the \acrshort{ps} and \acrshort{pl}. The tool \textit{PSU Configuration Generator} was developed to automate the creation of the binary configuration files, it could optionally also use the file \textit{base.xsa}.}
    \label{fig:createfilesext}
\end{figure}

Fig.~\ref{fig:createfilesext} shows the process to create all files used for the later stages of the boot process that can be fetched from a \acrshort{tftp} server in the network, including the binary configuration files to extend the configuration of the \acrshort{ps}. It can be seen that two paths are used to create the files. One uses PetaLinux tools to build the Linux kernel and the other path uses the custom tool PSU Configuration Generator to generate the two custom binary configuration files and to extract the bitfile for the \acrshort{pl}. In contrast to the files created for the early stages of the boot process, the files created here contain application-specific information and are thus mainly based on the *.xsa file exported from the Vivado project representing the complete configuration \textit{complete.xsa}. The information in \textit{base.xsa} can be used optionally to achieve a higher degree of automation.

The path in Fig.~\ref{fig:createfilesext} for building the Linux kernel uses solely the file \textit{complete.xsa} as input. This archive is used as the basis to set up a PetaLinux project. Afterwards, the PetaLinux framework fully automates the process of building the kernel. The tools provided by the PetaLinux software suite can be used to customize the kernel as usual.

To prevent redundant reconfigurations, the path in Fig.~\ref{fig:createfilesext} showing the usage of the PSU Configuration Generator could use both, \textit{complete.xsa} and \textit{base.xsa}. However, currently only the complete configuration is used as input. The redundant reconfigurations that occur because of this, have not caused any problems so far. However, the registers that can not be modified from U-Boot (see Table~\ref{tab:notAllowedReg}) must be declared in the JSON configuration files. Fig.~\ref{fig:createfilesext} also makes clear why it is efficient to use the PSU Configuration Generator to extract the bitfile for the \acrshort{pl} from \textit{complete.xsa}. This feature helps to have as many of the files that must be copied to the remote server ready at the same time and at the same location. Only the Linux kernel needs to be collected from a different location.

\section{Implementation and Testing}

The split boot mechanism as described here was developed and tested on a Trenz Electronic TE0803-03-4BE11-A MPSoC \acrfull{som} plugged onto a custom carrier board\,\cite{Ardila_2020} that included, among other things, an \acrshort{ssd}, two UART interfaces, and two network interfaces, one via \acrshort{sgmii} and one via \acrshort{rgmii}. On the software side, the development tools of the Xilinx toolset 2020.2 were used.

\begin{figure}[b!]
    \centering
    \includegraphics[width=3.4in]{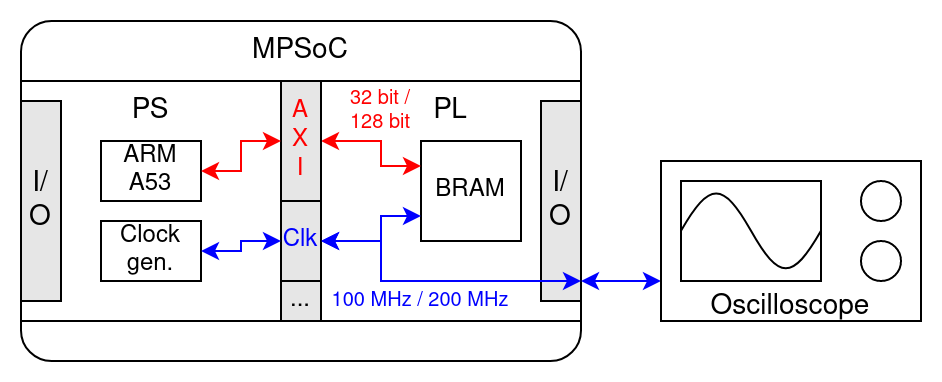}
    \caption{Setup used to test the reconfigurability of the AXI and clock interfaces between \acrshort{ps} and \acrshort{pl} by the split boot mechanism. After the initial configuration, both interfaces were enabled with 32-bit AXI width and 100 MHz clock. Later they were changed to 128-bit and 200 MHz.}
    \label{fig:measure}
\end{figure}

Because the split boot process in its most efficient implementation loads the configuration for the \acrshort{pl} from a network server, the ability to configure the interfaces between \acrshort{ps} and \acrshort{pl} at run time is of great interest. Two independent tests were run for validation. With the clocks generated in the \acrshort{ps} directly connected to \acrfull{mio} pins of the PL, the ability of activating the signal and changing the frequency was confirmed with an externally connected oscilloscope. The second test targets the \acrshort{axi} interfaces. A \acrshort{bram} \acrshort{ip} core instantiated in the \acrshort{pl} was used to confirm the possibility to activate them at run time and to change the width of the bus. Fig.~\ref{fig:measure} shows the setup used for both tests.

The reconfigurability of interfaces using \acrshort{serdes} was examined using the connected \acrshort{ssd}. \acrshort{serdes} interfaces are highlighted in particular here, because they are not only configured but also calibrated by the \acrshort{fsbl} and this calibration step was also relocated to U-Boot. After changing the configuration and perform the calibrating in U-Boot, read and write access to the \acrshort{ssd} from Linux was possible without any limitations. Another interface using \acrshort{serdes} is Ethernet via \acrshort{sgmii}. The Ethernet interface, however, needs to be configured in the \acrshort{fsbl} because it is used in the split boot mechanism. Thus, the only test possible was to use U-Boot to clear the respective configuration registers with zeros before restoring the configuration values. This test was also successful. After rebinding the Ethernet driver in U-Boot, the interface could be used normally. The same procedure was also successfully tested with the Ethernet interface based on \acrshort{rgmii} that consequently does not use \acrshort{serdes}. In addition, it was also tested whether the configuration of the \acrshort{mio} pins of the \acrshort{ps} can be changed. For this purpose two \acrshort{mio} pins were assigned to one of the \acrshort{uart}s in the \acrshort{ps} at run time. After that, the \acrshort{uart} could be used without restrictions for input and output. 

Aside from these tests aiming at the configurability of a single component, booting Linux on the \acrshort{mpsoc} after extending the configuration in U-Boot was used as a comprehensive test. This is possible because the majority of components in the \acrshort{ps} that are configured as part of the complete configuration are targeted and initialized by a Linux driver loaded during the kernel's boot process. Linux was able to boot on the reconfigured \acrshort{mpsoc} in the same way as if the \acrshort{ps} had been fully configured in the \acrshort{fsbl}. This supports the claim that after reconfiguration in U-Boot, the \acrshort{mpsoc} behaves exactly as if the configuration had been done completely in the \acrshort{fsbl}.

To investigate whether the isolation configured in U-Boot behaves the same way as if it had been configured in the \acrshort{fsbl}, two types of tests were run. The access to different regions in the address range of the DDR memory, separated by the isolation, was examined before and after the isolation was enabled. A similar access check was also performed for multiple registers belonging to different isolated components within the \acrshort{ps}. In both cases, the isolation behaved the same way as if it had been activated in the \acrshort{fsbl}. This outcome was expected because, despite the fact that the configuration of the isolation is handled in software, the actual separation of the \acrshort{ps} into multiple subsystems is enforced directly by hardware and thus unaffected by the order in which the software is executed\,\cite{xlnx_ug1085}.

\begin{figure}[tb!]
    \centering
    \includegraphics[width=2in]{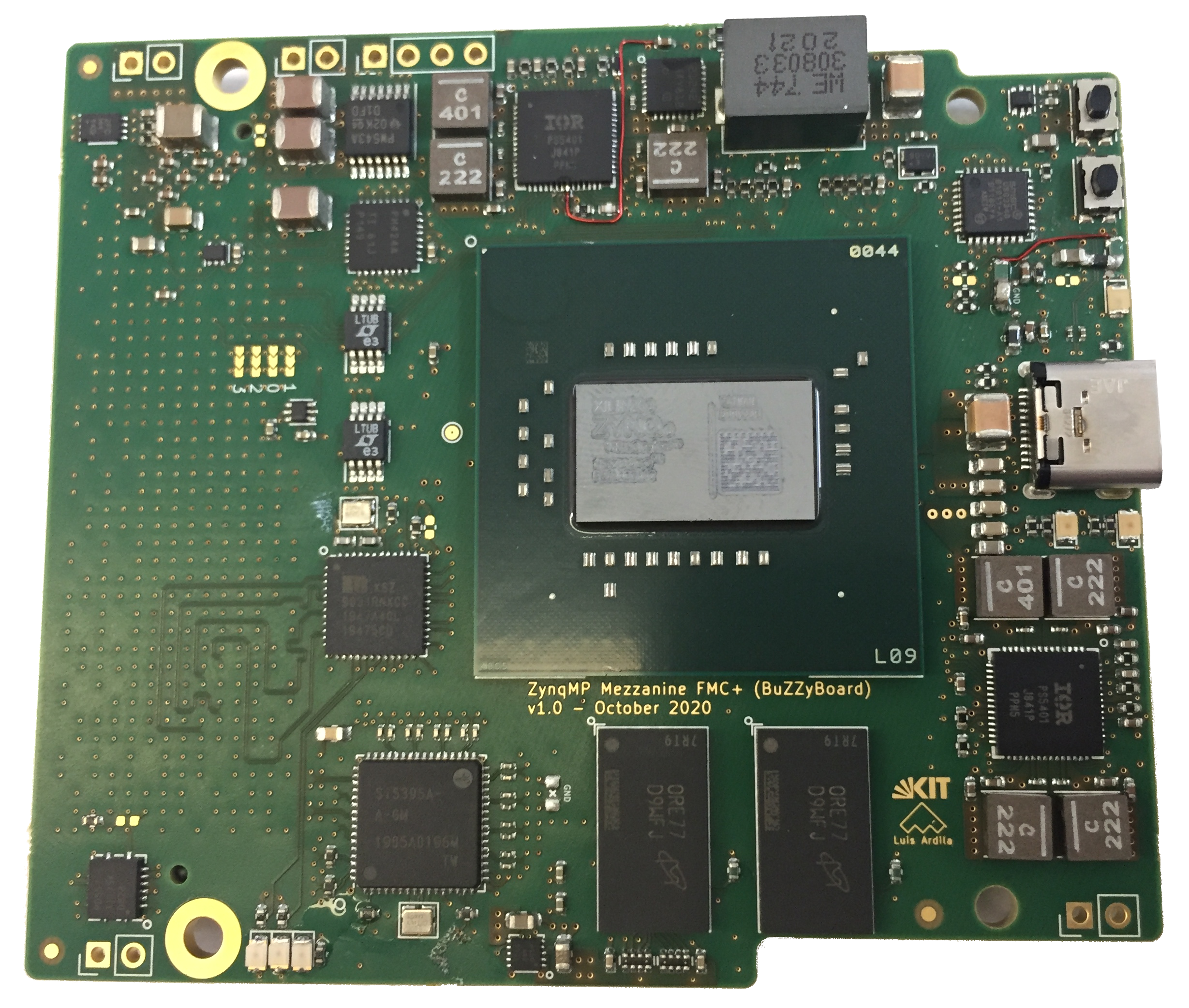}
    \caption{Custom Zynq Ultrascale+ MPSoC based FMC+ mezzanine board designed for slow control tasks.}
    \label{fig:buzzyboard}
\end{figure}

In addition to the Trenz Electronic MPSoC, split boot based on version 2020.2 of the Xilinx development tools was also implemented on a Xilinx ZCU102 evaluation board and on a custom \acrshort{zusp} based \acrshort{fmcp} mezzanine board, depicted in Fig.~\ref{fig:buzzyboard}\,\cite{2022JInst..17C3009M}. Furthermore it was implemented on a Xilinx Kria K26 \acrshort{som} plugged onto a KV260 development platform using version 2020.2.2 of the Xilinx toolset. Despite some minor changes to the patches required due to the different version of the toolset used for the Kria K26, the test results were identical. The implementation process on these different hardware platforms was also used to estimate the effort required to create all the projects and files needed for a new platform. Due to the two Vivado and PetaLinux projects used, the process takes longer than with the regular boot process, but the additional time required was typically well under an hour, especially when the patches for the version of the toolset used were already available.

\section{Conclusion}

The large number of hundreds of \acrshort{soc} devices used within the LHC upgrade creates significant challenges in their firmware deployment, maintenance, and accessibility. Booting from a singular source would be beneficial and would significantly ease maintenance. This functionality is supported by the modified boot process presented in this paper. The split boot process enables a clear separation by having all application-specific data on a remote server and just a generic base layer of software remaining on the local boot medium. The proposed workflow minimizes the overhead of implementing the modified boot process while relying on official Xilinx tools wherever possible. Split boot was implemented and tested on four different hardware platforms with two versions of the Xilinx development tools. Although the boot sequence is already fully functional, there is still room for improvement. A higher level of automation could be attained and will be addressed in future work.

\appendices

\end{document}